\documentclass[12pt]{iopart}

\usepackage{iopams}
\usepackage{graphicx}
\usepackage{color}
\usepackage{soul}
\usepackage{soul}

\begin{document}

\title{Spatial Current Patterns, Dephasing and Current Imaging in Graphene Nanoribbons}

\author{Joel Mabillard$^{1}$, Tankut Can$^{2}$ and Dirk K. Morr$^{1,2}$}

\address{$^{1}$ University of Illinois at Chicago, Chicago, IL 60607, USA\\
$^{2}$ Department of Physics and James Franck
Institute, University of Chicago, Chicago, IL 60637, USA}
\ead{\mailto{dkmorr@uic.edu}}
\begin{abstract}
Using the non-equilibrium Keldysh Green's function formalism, we investigate the local, non-equilibrium charge transport in graphene nanoribbons (GNRs). In particular, we demonstrate that the spatial current patterns associated with discrete transmission resonances sensitively depend on the GNRs' geometry, size, and aspect ratio, the location and number of leads, and the presence of dephasing. We identify a relation between the spatial form of the current patterns, and the number of degenerate energy states participating in the charge transport. Furthermore, we demonstrate a principle of superposition for the conductance and spatial current patterns in multiple-lead configurations. We demonstrate that scanning tunneling microscopy (STM) can be employed to image spatial current paths in GNR with atomic resolution, providing important insight into the form of local charge transport. Finally, we investigate the effects of dephasing on the spatial current patterns, and show that with decreasing dephasing time, the current patterns evolve smoothly from those of a ballistic quantum network to those of classical resistor network.

\end{abstract}



\maketitle

\section{Introduction}

Understanding charge transport in graphene nanoribbons (GNRs) has attracted significant interest in recent years \cite{Mun06,Cre07,Zar07,Li08,Mai09,Wak09,Dub09,Gon10,Tak10,Tak11,Sar11,Neto09}, in particular due to their potential application as integrated circuits \cite{Are07} and field-effect transistors \cite{Ouy06,Yan07,Chen07}, as bio-sensing devices \cite{Lu09,Pum11,Cho12}, and for DNA sequencing \cite{Sch10,Min11,Ven11,Saha12}. These unprecedented opportunities have been made possible by experimental advances in creating sub-10nm wide GNRs \cite{Li08a}, in engineering GNRs with specific electronic structures \cite{Han07}, in fabricating high purity samples \cite{Bol08}, and in designing artificial molecular graphene \cite{Gom12}. Moreover, due to their long mean-free path, GNRs are also an ideal system to explore the fundamental properties of out-of-equilibrium charge transport in nanoscopic quantum systems.

Previous theoretical studies investigating the transport properties of GNRs have predominantly focused on the bias dependent conductance \cite{Mun06,Cre07,Zar07,Li08,Mai09,Wak09,Dub09,Gon10,Tak11}. Spatial current patterns were investigated in the vicinity of the Fermi energy of half-filled GNRs in the wide-lead limit both for zero-magnetic field \cite{Zar07} and for non-zero magnetic fields and disorder \cite{Kum10}. Recently, however, it was argued \cite{Can12} that in nanoscopic networks with narrow constrictions and/or narrow leads, spatial current patterns emerge which are qualitatively different from those in the wide-lead limit. In particular, these spatial current patterns exhibit clear signatures of quantum behavior: they (a) possess coherent ``current riverbeds", i.e., spatial regions of large current density,  whose widths are of the order of the Fermi wave-length, and (b) are strongly dependent on boundary conditions, such as the position of the leads, the geometry and size of the network, as well as the gate voltage.
These results clearly suggest that understanding the quantum nature of local charge transport in GNRs, as reflected in the form of spatial current patterns, and its dependence on a GNR's geometry and size,  or the presence of dephasing, and developing a method to visualize it, is of utmost importance for the further development of graphene based nanoelectronics and DNA sequencing.

In this article, we address this open question by demonstrating how the spatial current patterns in GNRs are determined by the interplay between the GNRs' geometry, aspect ratio and size, by the location and number of leads, and the presence of dephasing. By using the non-equilibrium Keldysh Green's function formalism \cite{theory,Car71a} we find that the GNR's unconventional electronic structure \cite{Sar11} is reflected not only in the bias dependence of the conductance, but also in the form of the spatial current paths. In particular, we find that each of a GNR's discrete transmission resonances possesses a characteristic spatial current pattern (hence we refer to these resonances also as {\it current eigenmodes} in analogy to the equilibrium eigenmodes in the local density of states of nanoscopic systems). We demonstrate that the spatial form of these current patterns is determined by the number of degenerate states participating in the charge transport, which in turn allows us to predict how these current patterns evolve with increasing
size or aspect ratio of a GNR. We show that current patterns can include closed loops of circulating currents, as well as exhibit {\it backflow}, i.e., the flow of charge through certain links opposite to the direction of the net charge flow \cite{backflow}. We demonstrate that the spatial current paths are qualitatively different between leads attached to the zig-zag or armchair edges of the GNR and explore the form of charge transport parallel and perpendicular to edge states. Moreover, we show  that while current patterns in certain 4-lead configurations can arise from the superposition of two 2-lead configurations, the conductance does not necessarily obey the superposition principle. Furthermore, we demonstrate that the quantum behavior of local charge transport can be visualized using scanning tunneling microscopy (STM) \cite{Can13}: it represents an essentially non-intrusive method to image spatial current paths in GNR with atomic resolution, providing important insight into the form of local charge transport. Finally, we investigate more realistic models including the effects of dephasing and next-nearest-neighbor hopping, and show that with decreasing dephasing time, the current patterns evolve smoothly from those of a ballistic quantum network to those of a classical resistor network. These results provide important insight into the fundamental aspects of charge transport in nanoscopic graphene lattices.

%
%

\section{Theoretical Model}
\label{sec:theory}

\begin{figure}
\begin{center}
\includegraphics[width=10cm,clip]{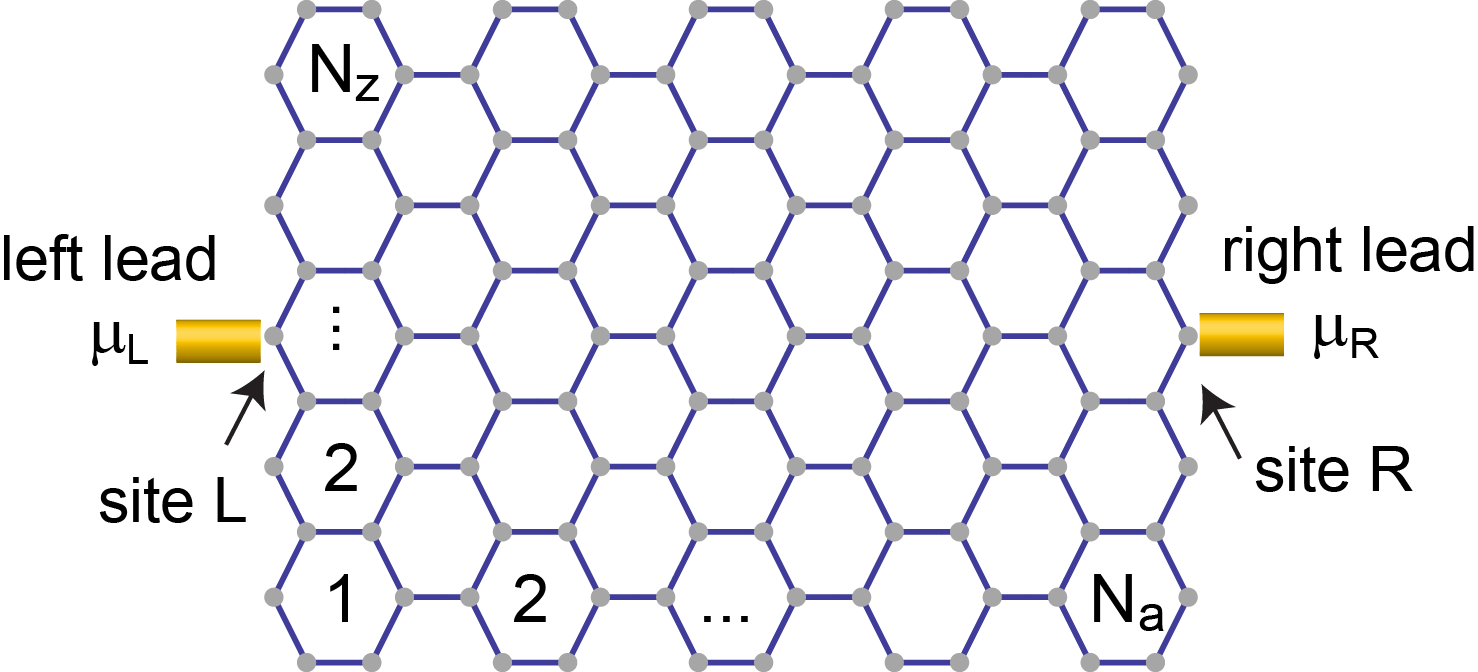} \caption{Schematic picture of an $(N_a \times N_z)$ graphene nanoribbon (GNR). The leads are connected to GNR sites ${\bf L}$ and ${\bf R}$.} \label{fig:NGL}
\end{center}
\end{figure}

To study electron transport in a graphene nanostructure, we consider two one-dimensional leads coupled to a finite $(N_a\times N_z)$ honeycomb lattice with $N_{a}$ hexagonal cells in the armchair direction, and $N_{z}$ cells in the zig-zag direction, shown schematically in figure~\ref{fig:NGL}.
%
%
The graphene nanoribbon is described by the Hamiltonian \cite{Oni08}
\begin{equation}
{\cal H}_{GNR} =   -  \sum_{{\bf r},{\bf
r}^\prime,\sigma} t_{\bf r,r^\prime} \; c^\dagger_{{\bf r},\sigma} c_{{\bf r}^\prime,\sigma}
 + \sum_{\bf r} \omega_0 a_{\bf r}^\dagger a_{\bf r} + g \sum_{\bf r, \sigma} (a_{\bf r}^\dagger + a_{\bf r}) c^\dagger_{{\bf r},\sigma} c_{{\bf r},\sigma}
\label{eq:Hamiltonian}
\end{equation}
where $-t_{\bf r,r^\prime}$ is the electronic hopping matrix element between sites $\bf r,r^\prime$ in the GNR, and $c^\dagger_{{\bf r},\sigma}$ creates an electron with spin $\sigma$ at site ${\bf r}$. In what follows, we take only the nearest neighbor hopping element, $t$, to be non-zero; the effects of a next-nearest neighbor hopping $t^\prime$ will be discussed in Sec.~\ref{sec:NNN}. In order to account for the effects of dephasing, we assume that the electrons interact locally (with coupling constant $g$) with a phonon mode of energy $\omega_0$, as described by the last two terms in equation (\ref{eq:Hamiltonian}), where $a_{\bf r}^\dagger$ creates a phonon at site ${\bf r}$. The coupling between the GNR and the leads is described by the Hamiltonian
\begin{equation}
{\cal H}_c = - t_h \sum_{{\bf r},{\bf l},\sigma} \; \left( c^\dagger_{{\bf r},\sigma} d_{{\bf l}, \sigma} + h.c. \right)
\end{equation}
where the primed sum runs over all sites ${\bf r}$ and ${\bf l}$ in the GNR and leads, respectively, that are coupled by a hopping element $t_h$, and $d^\dagger_{{\bf l}, \sigma}$ creates an electron with spin $\sigma$ at site {\bf l} in the leads. Below, we assume that each of the two leads is coupled to a single GNR site only, labeled ${\bf L}$ and ${\bf R}$ (see figure~\ref{fig:NGL}).
Therefore, the only relevant property of the leads entering our calculations is the local Green's function at the lead sites that are coupled to the GNR. Finally, for the purpose of imaging spatial current patterns in the GNR \cite{Can13}, we consider the tunneling of electrons from an STM tip to a single site {\bf T} in the GNR, a process which is described by the Hamiltonian
\begin{equation}
{\cal H}_{tip} = - t_T  \; \left( c^\dagger_{{\bf T},\sigma} f_{\sigma} + H.c.\right)
\label{eq:tunnel}
\end{equation}
where $f_{\sigma}$ annihilates an electron with spin $\sigma$ in the STM tip.

The spatial current patterns in a GNR are obtained by computing the current, $I_{\bf r  r^\prime}$, between adjacent sites ${\bf r}$,${\bf
r}^\prime$ in the GNR. This current is induced by different chemical potentials, $\mu_{L,R}$ in the left and right leads, and given by \cite{Car71a}
\begin{equation}
I_{\bf r  r^\prime}=-2 \frac{e}{\hbar} \; t
\intop_{-\infty}^{+\infty}\frac{d\omega}{2\pi}{\rm Re} \left[
G^K_{\bf r  r^\prime}(\omega)\right] \ , \label{eq:Current}
\end{equation}
where $G^K_{\bf r  r^\prime}$ is the full Keldysh Green's function between sites  ${\bf r}$ and ${\bf
r}^\prime$ which accounts for the electronic hopping within the GNR and between the GNR and the leads, as well as the electron-phonon interaction. We next introduce a matrix notation such that the Keldysh Green's function matrix is given by ${\hat G}^{K}$, and its $(ij)$ element, ${\hat G}^{K}_{ij}$, is the Keldysh Green's function between sites in the GNR denoted by $i$ and $j$. In the absence of an electron-phonon interaction, one finds
$ \hat{G}^{K} =
\left(1-\hat{g}^{r}\hat{t}\right)^{-1}\hat{g}^{K}\left(1-\hat{t}\hat{g}^{a}\right)^{-1}$
with $
{\hat g}^{K}(\omega) = 2i \left[1-2 {\hat n_F}(\omega) \right] {\rm Im}
\left[ {\hat g}^r(\omega) \right]$. Here, $\hat{g}^{r,a,K}$ are the diagonal retarded,
advanced and Keldysh Green's function matrices, respectively, containing the Green's functions of the
decoupled ($t,t_h=0$) GNR sites and leads: for the GNR sites, we have $g^{r}=1/(\omega +i\delta)$ with $\delta=0^+$ while for the leads, we assume the wide-band limit and take $g^{r} = - i \pi t^{-1} $, yielding a constant density of states $N_{L} = t^{-1}$ in the leads. Finally, ${\hat n_F}$ is a diagonal matrix containing the
Fermi-distribution functions and $\hat{t}$ is the symmetric hopping matrix. Moreover, the local density of states at site ${\bf r}$ in the GNR is obtained via $N({\bf r},\omega) = - {\rm Im}G^r_{\bf r  r}(\omega)/\pi$. For the results shown below, we have used for numerical convenience $\delta = k_{B} T = 10^{-5} t$, $\Delta \mu = \mu_{L} - \mu_{R} = 2 \times 10^{-5}t$ and $t_h=0.1t$, unless otherwise specified. All spatial currents, $I_{\bf r r^\prime}$, presented below are normalized to the largest current in the GNR.

To account for the effects of the electron-phonon interaction for general $g, \omega_0$ and $T$ is computationally demanding and beyond the scope of this article. However, since we are mainly interested in the effects of dephasing on the spatial current patterns, we can consider the high-temperature approximation introduced in Ref.~\cite{Bih05}. In this limit with $k_B T \gg \omega_0$, implying a large thermal population of the phonon mode, the calculation of the electronic self-energy correction is greatly simplified and computationally possible even for larger GNR sizes. Retaining in the Dyson equation only those terms that contain a factor of $n_B(\omega_0) \gg 1$, the electronic self-energy in the self-consistent Born approximation is given by
\begin{equation}
\Sigma_{ii}^{\alpha}(\omega) = \frac{i g^{2}}{2} \int \frac{d\nu}{2\pi} D^{K}(\nu) G_{ii}^{\alpha}( \omega - \nu)
\end{equation}
where $i$ denotes a site in the GNR and $\alpha = K, r, a$. Moreover,
\begin{equation}
D^{K} = 2 i \pi \left[ 1 + 2 n_{B}^{ph}(\omega)\right] \left[ \delta(\omega + \omega_{0}) - \delta(\omega - \omega_{0})\right]
\end{equation}
is the Keldysh phonon Green's function, which we assume to remain unchanged by the electron-phonon interaction, and $n^{ph}_{B}(\omega)$ is the phonon Bose distribution function (we assume that the phonons remain in thermal equilibrium). Note that due to the coupling to local phonon modes, the electronic self-energy is entirely local. A further simplification is achieved by considering the limit $\omega_{0} \rightarrow 0$ such that the self-energy, to leading order in $k_{B}T/\omega_{0}$, is given by
\begin{equation}
\Sigma_{ii}^{\alpha}(\omega) =  2 g^{2} \frac{k_{B}T}{\omega_0} G_{ii}^{\alpha}(\omega) \equiv \gamma G_{ii}^{\alpha}(\omega) \ .
\label{eq:Sigma}
\end{equation}
As shown before \cite{Can12,Bih05}, the solution of the Dyson equation for the retarded Green's function is then given by
\begin{equation}
\hat{G}^{r} = \left[ 1 - \hat{g}^{r} \left( \hat{V} + \gamma {\tilde D} {\hat G}^r \right) \right]^{-1} \hat{g}^{r}
\label{eq:SE2}
\end{equation}
where $\tilde{D}$ is a superoperator introduced in Ref.~\cite{Bih05} which, when operating on a Green's function matrix, returns the same matrix with all elements set to zero except for the diagonal elements in the matrix that correspond to sites in the GNR, e.g.,
\begin{equation}
[ \tilde{D} \hat{G}^{\alpha}]_{ij} = \cases{G_{ij}^{\alpha}\delta_{ij} &if $i$ is a site in the GNR\\
0& otherwise \ . \\}
\end{equation}

After self-consistently solving equation (\ref{eq:SE2}) for $\hat{G}^{r}$, $\hat{G}^{K}$ can be obtained in a closed expression via
$\hat{G}^{K} =  {\hat G}^{r} {\tilde \Sigma} {\hat G}^{a}$
where the diagonal matrix ${\tilde \Sigma}$ is defined via
\begin{equation}
{\tilde \Sigma}_{ ll}=  \left [ \left(1 - \gamma {\hat Q} \right)^{-1} \boldsymbol{\lambda} \right]_{ l} .
\end{equation}
Here, the vector $\boldsymbol{\lambda}$ has components $\boldsymbol{ \lambda}_{m} = {\hat \Lambda}_{ m m}$ with $\hat{\Lambda} = \left(\hat{g}^{r} \right)^{-1} \hat{g}^{K} \left( \hat{g}^{a} \right)^{-1}$ being a diagonal matrix whose only non-zero elements $\hat{\Lambda}_{ii}$ are those where $i$ denotes one of the two leads. The matrix ${\hat Q}$ contains the elements
\begin{equation}
{\hat Q}_{lm}=\cases{\left|G_{ lm}^{r}\right|^{2}   &if ${ l}$ is a site in the GNR\\
 0&  otherwise\\}
\end{equation}

\section{Results}

\subsection{Current Patterns in the $ (11 \times 5)$ GNR}

%
%
\begin{figure} \begin{center}
\includegraphics[width=15.cm]{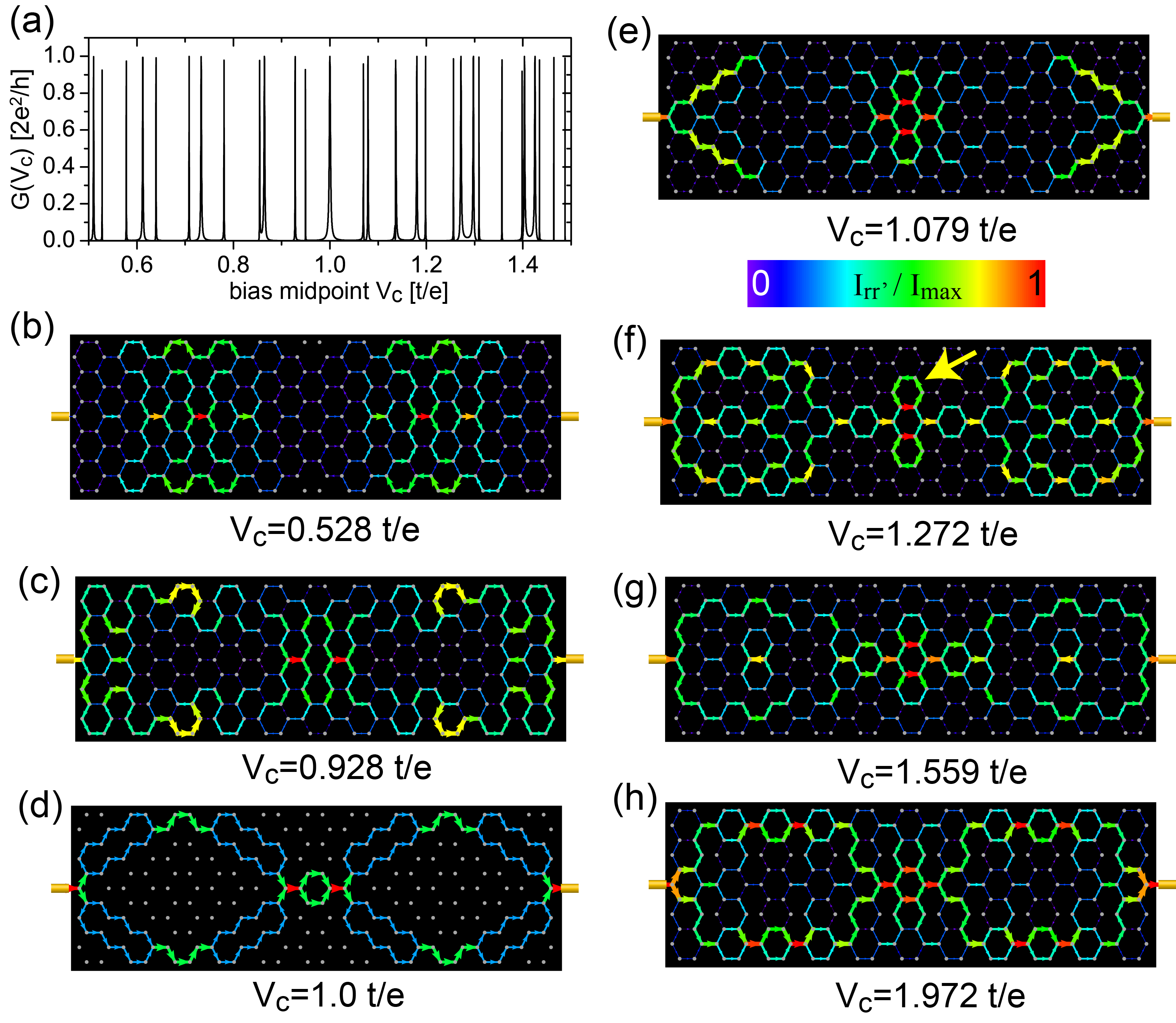} \caption{ (a) Conductance of an $ (11 \times 5)$ GNR, and (b) - (h) spatial current patterns for different $V_c$. All spatial currents, $I_{\bf r r^\prime}$, are normalized to the largest current in the GNR. Yellow arrow in (f) indicates a circulating current loop. The leads are connected to GNR sites ${\bf L}$ and ${\bf R}$.} \label{fig:NGL11x5}
\end{center}
\end{figure}
We begin by discussing charge transport, and in particular the relation between the conductance and spatial current patterns, in an $(11 \times 5)$ GNR in the absence of the electron-phonon interaction.
In figure~\ref{fig:NGL11x5}(a), we present the GNR's total conductance $G(V_c) = I(V_{c})/\Delta V$ in the limit $\Delta V =(\mu_L-\mu_R)/e \rightarrow 0$ which is given by
\begin{equation}
G(V_c) = 4\pi \frac{e^{2}}{\hbar} t_{h}^4  N_0^2 |G^r_{{\bf L,R}}(\mu_c)|^2
\end{equation}
where $V_c=\mu_c/e$ is the bias midpoint with $ \mu_c=(\mu_L+\mu_R)/2$, $N_{0}$ is the leads' local density of states, and $G^r_{{\bf L,R}}$ is the non-local Green's function  between sites ${\bf L}$ and ${\bf R}$ where the current enters and exits the GNR (see figure~\ref{fig:NGL}), respectively \cite{Car71a}.
Due to its finite size, the GNR possesses discrete energy levels \cite{Oni08} which for $t_h=0$ are given by
\begin{equation}
\left[ E_{j}^{\pm}/t \right]^2 =  1 \pm 4\cos\frac{\pi j}{2 (N_{z} + 1)} \cos \frac{\kappa_{j}^{\pm}}{2} + 4 \cos^{2}\frac{\pi j}{2(N_{z}+1)}
 \end{equation}
where the  $\kappa_{j}^{\pm}$ are solutions of
\begin{equation}
 \frac{\sin \kappa_{j}^{\pm} N_{a}}{\sin \kappa_{j}^{\pm}(N_{a} + 1/2)} = \mp 2 \cos \frac{\pi j}{2 (N_{z}+1)} \ .
\end{equation}
As a result, the conductance exhibits discrete transmission resonances whenever $\mu_c$ equals the energy a state whose wavefunction does not vanish at the site that the leads are coupled to. For each of these resonances, the current flowing through the GNR exhibits a different spatial pattern (the resonances are therefore also referred to as current eigenmodes \cite{Can12}, in analogy to eigenmodes in the density of states). The spatial current patterns [as obtained from equation (\ref{eq:Current})] for a number of current eigenmodes are shown in figures~\ref{fig:NGL11x5}(b) - (h) [due to the particle-hole symmetry of the GNR's electronic structure, the conductance is symmetric around $V_c=0$ and the eigenmodes at $\pm V_c$ exhibit identical current patterns; we therefore restrict our discussion below to the case $V_c \geq 0$]. Note that these eigenmodes can in general be accessed experimentally via gating of the GNR. Similar to prior results \cite{Can12}, we find that there exists current eigemodes which exhibit circulating current loops [as indicated by a yellow arrow in figure~\ref{fig:NGL11x5}(f)] or which possess a net current through the GNR which is significantly smaller than some of the currents within the GNR [see figure~\ref{fig:NGL11x5}(b)]. Of particular interest is the well-ordered current pattern that occurs for $V_c= t/e$ shown in figure~\ref{fig:NGL11x5}(d). To understand the spatial form of this current pattern, we note that charge transport for $V_c= t/e$ with leads attached to the zig-zag edge involves a $5-$fold degenerate $E=t$ state (not including $N_a=5$ states with vanishing amplitude along the zig-zag edge \cite{Oni08}), whose wave-vectors are shown in figure~\ref{fig:x5}(a) (solid and open circles); for comparison, we also present the $E=t$ equal energy contour for an infinitely large graphene sheet (solid line). Only three of these 5 states [green circles in figure~\ref{fig:x5}(a)] possess a non-zero wavefunction at the GNR sites ${\bf L}$ and ${\bf R}$ that the leads couple to in figure \ref{fig:NGL11x5} and thus contribute to the charge transport at $V_c= t/e$.  The Fermi velocity at $E=t$ is given by ${\bf v}_{F} = \partial E/\partial {\bf k}= v_{F}\left( \sqrt{3}/2, 1/2\right)$ [see figure~\ref{fig:x5}(a)] and thus parallel to the primitive lattice vectors of one of the triangular sublattices of graphene. As follows from figure~\ref{fig:NGL11x5}(d), the current propagates generally along the same direction. Thus, the spatial current pattern is reminiscent of the motion of a ballistic particle that is specularly reflected off the walls of the GNR. We find that this relation between the spatial current pattern and ${\bf v}_{F}$ holds whenever there are degenerate states with the same ${\bf v}_{F}$ participating in the charge transport. The relation between the spatial current pattern for $V_c=t/e$, the aspect ratio of the GNR, and the degeneracy of the $E=t$ state will be discussed in more detail in section~\ref{sec:aspect}. It is noteworthy that, in the vicinity of $V_c=t/e$, the spatial form of the current patterns evolves rapidly with increasing $V_c$ [see figures~\ref{fig:NGL11x5}(c) - (e)] while for larger values of  $V_c$, the spatial form evolves more slowly [see figures~\ref{fig:NGL11x5}(f) - (h)].

Finally, we note that for finite-sized GNRs, there exist two states near zero energy at $\pm E_l$, that are delocalized along the zig-zag edge, but localized in the armchair direction \cite{Oni08}. For $V_c = \pm E_l/e$, the GNR therefore exhibits strongly anisotropic transport properties in the armchair and zig-zag directions, as discussed in more detail in section~\ref{sec:edge}.

\subsection{Spatial current patterns, aspect ratios, and the degeneracy of the $E=t$ state}
\label{sec:aspect}

%
%
\begin{figure} \begin{center}
\includegraphics[width=16cm,clip]
{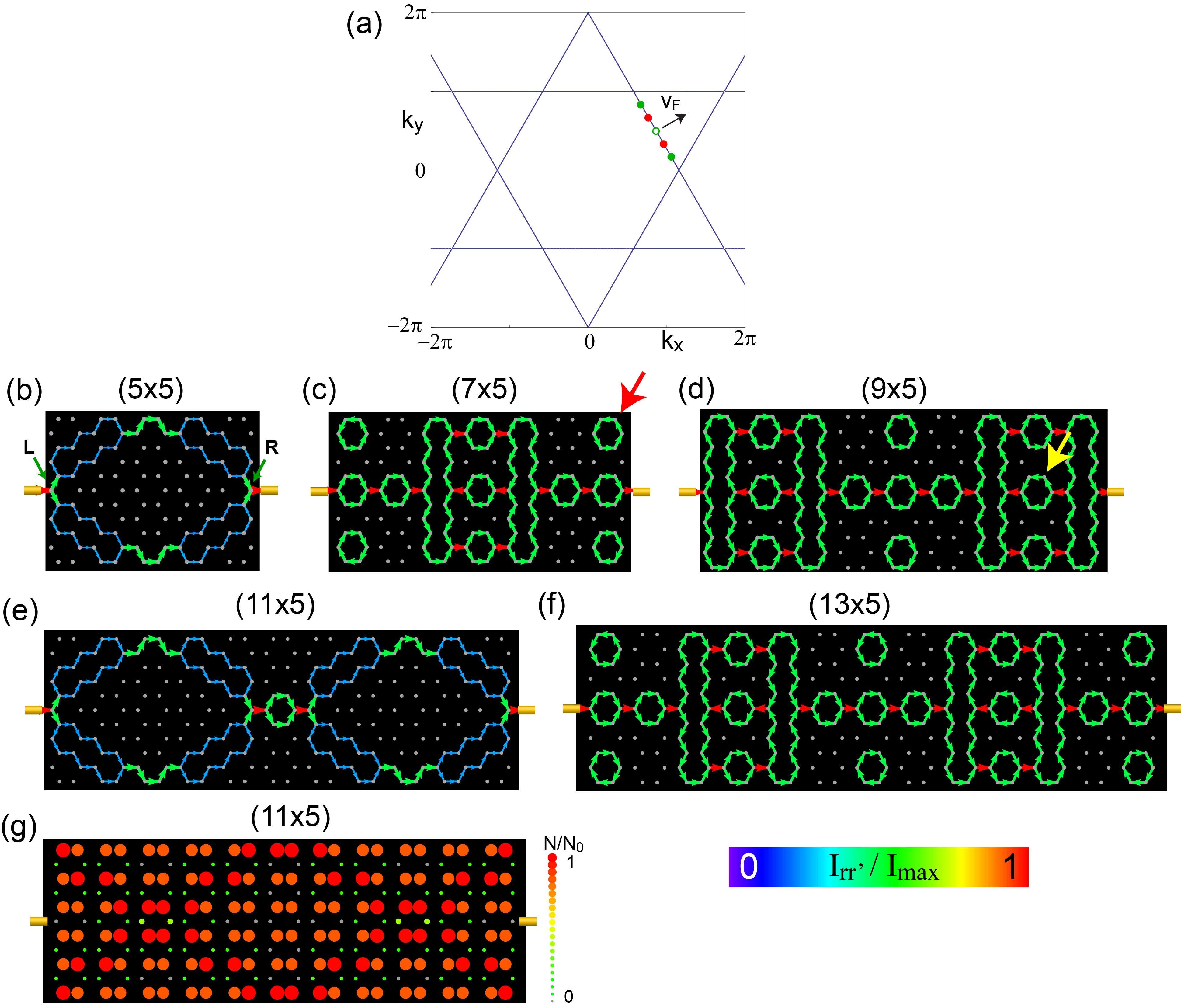} \caption{(a) Wave-vectors of the 5 degenerate $E=t$ states (closed and open circles) in an $(11 \times 5)$ GNR together with the $E=t$ equal energy contour in the Brillouin zone of an infinite graphene sheet. The wave-vectors of the three states possessing a non-zero wavefunction at site ${\bf L}$ and ${\bf R}$ are shown in green. (b) - (f) Evolution of the spatial current patterns for $V_c=t/e$ in $(N_a \times 5)$ GNRs with increasing $N_a$. (g) Contour plot of the normalized density of states of the $(11 \times 5)$ GNR at $E=t$.} \label{fig:x5}
\end{center}
\end{figure}
In the ballistic limit, spatial current patterns in nanoscale systems sensitively depend on boundary conditions such as the location of the leads or the geometry or aspect ratio of the system \cite{Can12}. To investigate this issue in GNRs, we next study the dependence of spatial current patterns on the aspect ratio, $N_a/N_z$ of the GNR, as well as the location of the leads. To exemplify this dependence, we consider the current eigenmode at $V_c=t/e$, since it does not only exhibit a spatially highly ordered structure, but also occurs in all GNRs considered below. In figure~\ref{fig:x5}, we present the evolution of the spatial current pattern with increasing $N_a$ in $(N_a \times 5)$ GNRs for $V_c=t/e$. We note for the discussion below that in all of these GNRs, there exist $N_a$ states at $E=t$ whose wavefunction vanishes at the sites that the leads are attached to \cite{Oni08}; those states are therefore irrelevant for the purpose of the charge transport considered here and will not be discussed below.

The spatial current patterns in the GNRs exhibit two types of characteristic forms. In particular, in GNRs that satisfy $N_{a} + 1= p(N_{z}+1)$ with integer $p$, the current follows the direction of the Fermi velocity, ${\bf v}_{F}$, being specularly reflected off the walls of the GNR, as shown in figures~\ref{fig:x5}(b) and (e). In these GNRs, there are $3$ states at $E=t$ whose wavefunctions do not vanish at sites ${\bf L}$ and ${\bf R}$; their wave-vectors are shown in figure~\ref{fig:x5}(a) (green circles), together with the $E=t$ equal-energy contour of an infinite graphene sheet. With increasing $p$, certain elements of the spatial current pattern are repeated [for example, the current pattern in figure~\ref{fig:x5}(e) with $p=2$ consists of two copies of the current pattern of figure~\ref{fig:x5}(b) plus an additional current loop connecting these two elements]. In contrast, the GNRs shown in figure~\ref{fig:x5}(c),(d), and (f) do not satisfy the above requirement, possessing only a single $E=t$ state [whose wave-vector is shown as the open circle in figure~\ref{fig:x5}(a)]. Consequently, these structures exhibit current patterns that are significantly more complex and possess a series of traits characteristic of quantum mechanical charge transport in nanostructures \cite{Can12}. In particular, these current patterns exhibit circulating current loops that are isolated [see red arrow in figure~\ref{fig:x5}(c)] and thus do not contribute to the net current through the GNR. Such current loops give rise to magnetic fields that could potentially be detected experimentally: we estimate that the magnetic field at the center of the current loop indicated by an arrow in figure \ref{fig:x5}(c) is approximately $5.5 \times 10^{-6}$T. In addition, these current patterns exhibit backflow: while a net current flows from the left to the right ($\mu_L > \mu_R$), there exist certain bonds in the GNR where the current flows from right to left (and thus opposite to the applied bias), such as the ones indicated by a yellow arrow in figure~\ref{fig:x5}(d). We find that circulating current loops and backflow occurs in all GNRs in which the current pattern does not simply reflect the direction of the Fermi velocity, ${\bf v}_{F}$, as is the case in figures~\ref{fig:x5}(b) and (e). Finally, in figure~\ref{fig:x5}(g) we present the local density of states for the $(11 \times 5)$ GNR (with leads attached) at $E=t$, which has no resemblance to the form of the spatial current pattern shown in figure~\ref{fig:x5}(e), a conclusion that was previously also drawn for nanoscale systems with a square lattice geometry \cite{Can12}. This result stands in contrast to earlier findings relating either the spatial current pattern at $V_c$ directly to that of the modulus of the wave-function, $|\Psi({\bf r},E=eV_c)|$ \cite{Tod99} (which possesses the same spatial structure as the density of states, $N({\bf r},E)$, since $N({\bf r},E) \sim |\Psi({\bf r},E=eV_c)|^2$), or the spatial flow of electrons to the spatial form of $|\Psi({\bf r},E=eV_c)|^2$ \cite{Top00_2}. More generally, our results demonstrate that conductance variations measured in SPM experiments cannot simultaneously reflect the spatial flow of electrons \cite{Top00} and the spatial form of $|\Psi({\bf r},E=eV_c)|^2$ \cite{Mar07}.

%
%
\begin{figure*} \begin{center}
\includegraphics[width=12.5cm,clip]
{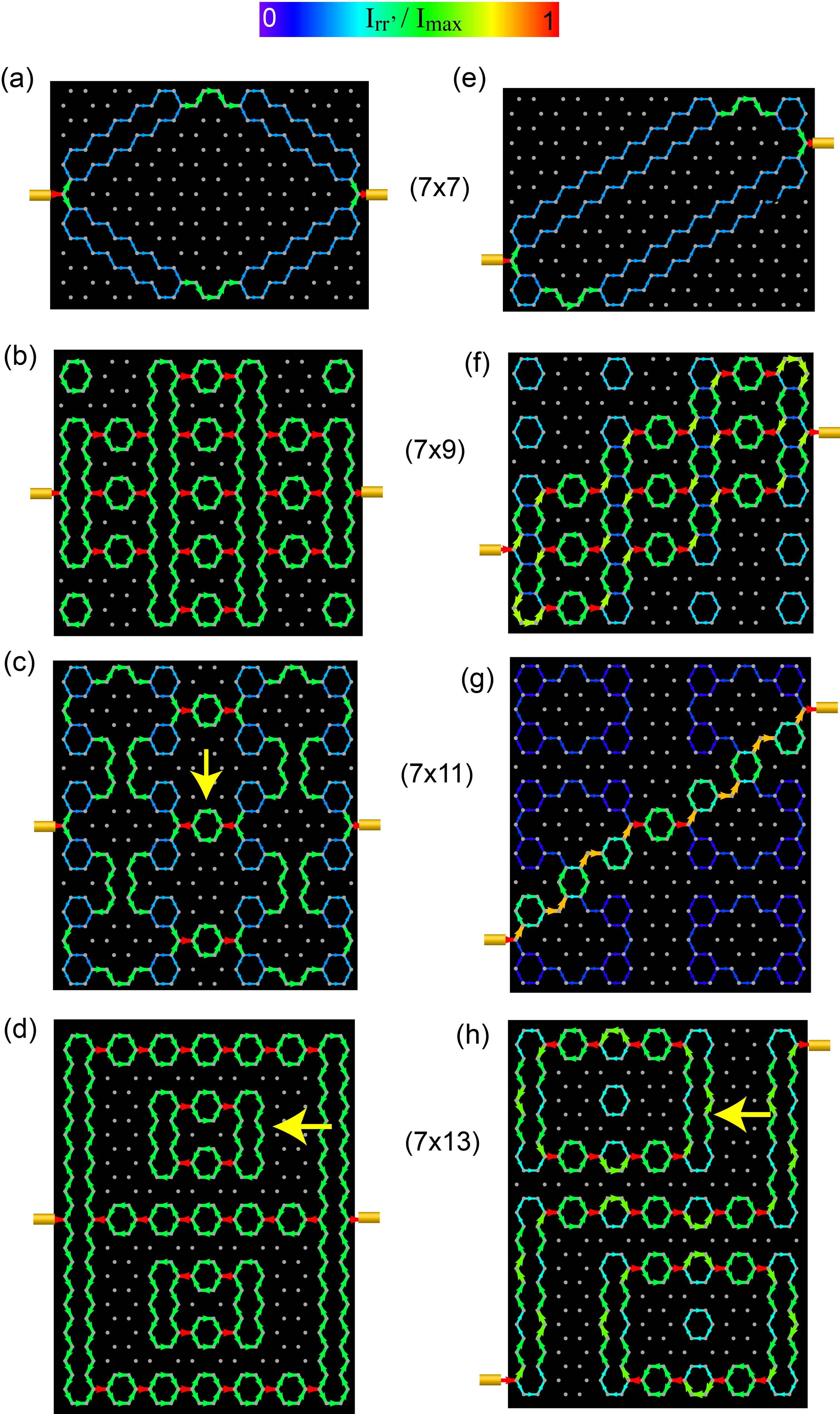} \caption{Evolution of the spatial current patterns for $(7 \times N_z)$ GNRs with increasing $N_z$ with (a) - (d) leads attached to the middle of the GNR, and (e) - (h) leads symmetrically displaced off-center.} \label{fig:7x}
\end{center}
\end{figure*}
Different types of current patterns emerge when the size of the GNR is increased along the zig-zag direction, as shown in figures~\ref{fig:7x}(a) - (d). Here, we present the evolution of the current patterns  with increasing $N_z$ in $(7 \times N_z)$ GNRs at $V_c=t/e$. The current patterns for the $(7 \times 7)$ GNR (which possesses a 7-fold degenerate state at $E=t$, with 4 of these states possessing a non-zero wavefunction at GNR sites ${\bf L}$ and ${\bf R}$) is similar to that of the  $(5 \times 5)$ or $(11 \times 5)$ GNRs, with the current propagating predominantly along the direction of the Fermi velocity. With increasing $N_z$, the current patterns become more complex, exhibit circulating current loops of various sizes [see yellow arrows in figures~\ref{fig:7x}(d) and (h)], and current backflow [see yellow arrow in figure~\ref{fig:7x}(c)]. We can identify an interesting relation between the number of degenerate states at $E=t$, and the form of the current pattern: while the $(7 \times 9)$ and $(7 \times 13)$ GNRs possess only a single state at $E=t$, and their spatial current patterns exhibit some similar subpatterns, the $(7 \times 11)$ GNR possesses three degenerate states (two of which possess a non-zero wavefunction at sites ${\bf L}$ and ${\bf R}$), resulting in a current pattern that is qualitatively different from that of the $(7 \times 9)$ and $(7 \times 13)$ GNRs. All of these current eigenmodes exhibit a different response to (symmetric) changes in the location of the leads, as shown in figures~\ref{fig:7x}(e)-(h). For the $(7 \times 7)$ GNR, the current pattern simply deforms, but the current still propagates along the direction of the Fermi velocity. In contrast, for the $(7 \times 9)$, $(7 \times 11)$ and $(7 \times 13)$ GNRs, the current pattern undergoes a qualitative change when the leads are symmetrically displaced. This reiterates the fact that charge transport through confined systems in the quantum regime is highly sensitive to boundary conditions, and the spatial form of the resulting coherent current eigenmode need not necessarily conform to expectations based on \({\bf v}_{F}\).

%
%
\begin{figure}
\begin{center}
\includegraphics[width=14cm,clip]
{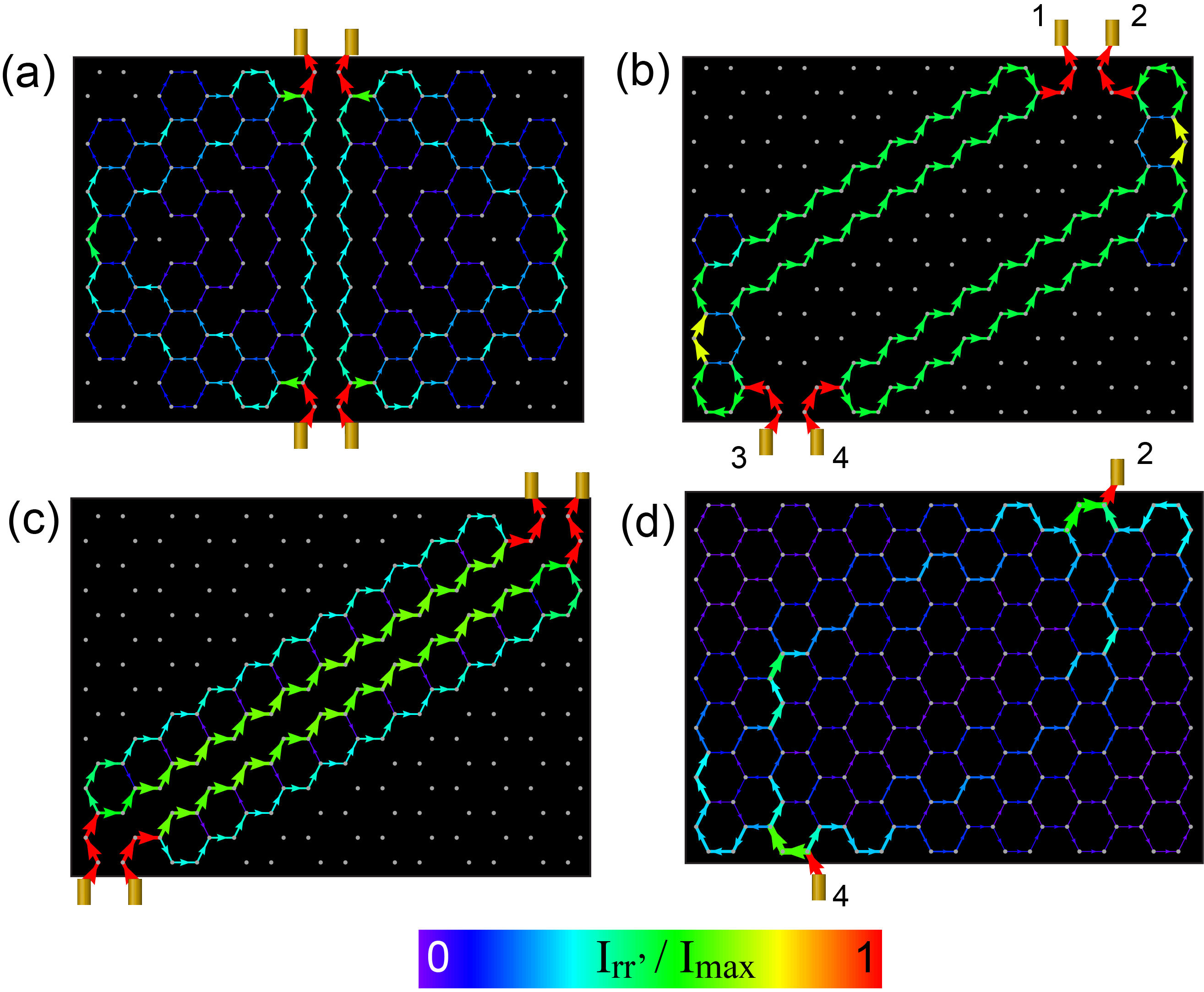} \caption{Current patterns in an $(7 \times 7)$ GNR with leads attached at different locations along the armchair edges.} \label{fig:arm}
\end{center}
\end{figure}
The question naturally arises whether the same type of spatial current patterns emerge when the leads are attached to the armchair edges. To investigate this question, we consider a $(7 \times 7)$ GNR [the same GNR as shown in figure~\ref{fig:7x}(a)] and present in figure~\ref{fig:arm} the current patterns at $V_c=t/e$ for different locations of 4 leads attached to the armchair edges. When the leads are attached to the center sites, as shown in figure~\ref{fig:arm}(a), the dominant contribution to the current flows along a line directly connecting the leads, rather than following the direction of the Fermi velocity. However, when the leads are displaced symmetrically from these high symmetry sites as shown in figures~\ref{fig:arm}(b) and (c), the current flows again primarily along the direction of the Fermi velocity. It is interesting to note that in all three cases [figures~\ref{fig:arm}(a)-(c)], the current exhibits two disjoint current paths, connecting either the left (leads 1 and 3) or right (leads 2 and 4) pair of leads. When one of these pairs of leads is removed (and the spatial symmetry of the system is thus broken), as shown in figures~\ref{fig:arm}(d) where only leads 2 and 4 remain, the current pattern becomes rather diffuse (a very similar current pattern is found when only leads 2 and 3 remain). This not only demonstrates that the symmetry of the armchair edges requires two leads to be attached to either armchair edge for a well-defined current pattern to emerge, but also that quantum interference effects between all four leads are crucial for creating the well-ordered current patterns shown in figures~\ref{fig:arm}(b) and (c).

\subsection{Current Through Localized Edge States}
\label{sec:edge}

\begin{figure}
\begin{center}
\includegraphics[width=12.0cm]{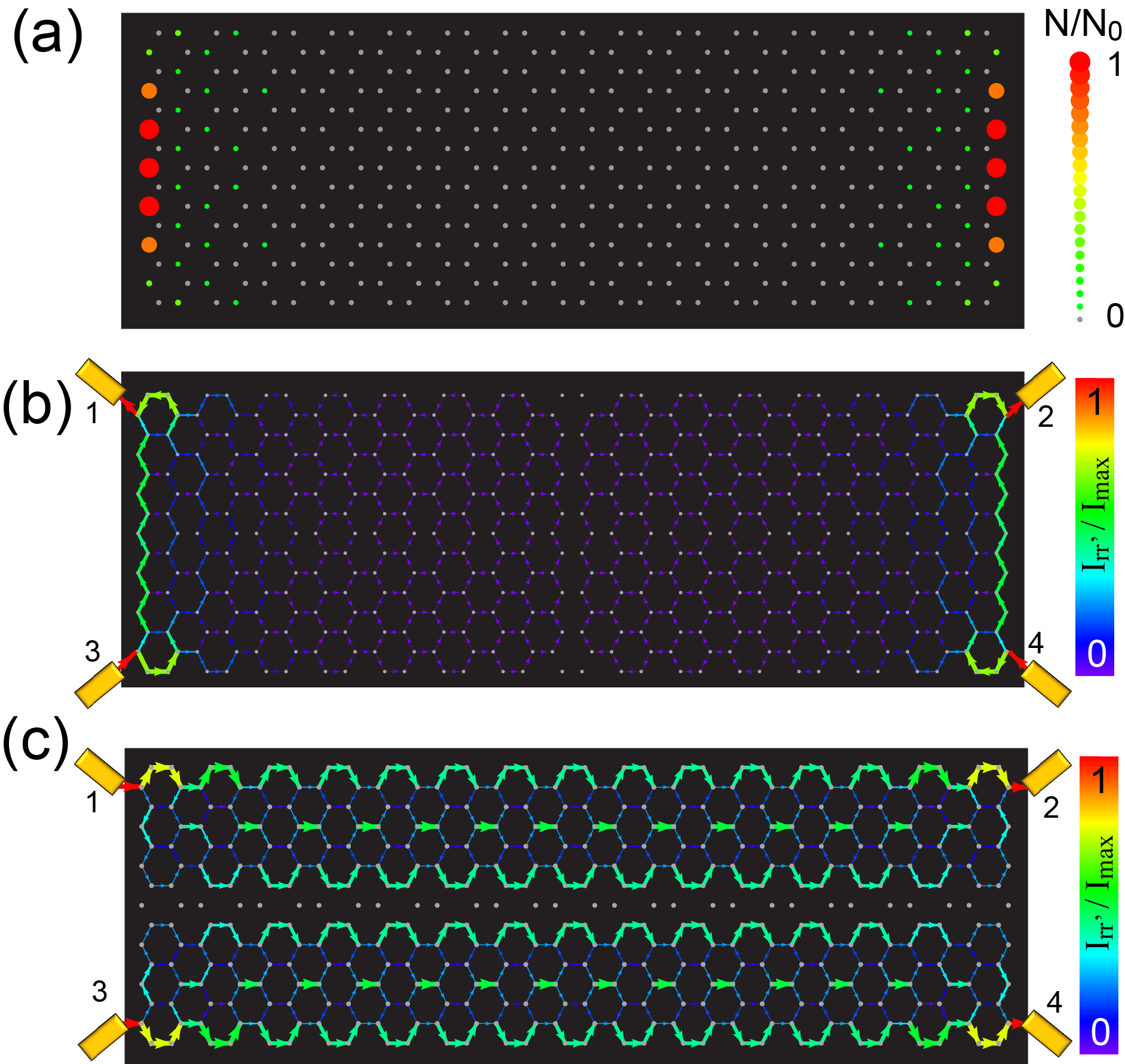} \caption{(a) Spatial plot of the local density of states at $E=E_l=4.6 \times 10^{-13} t$ demonstrating the existence of an edge state that is delocalized along the zig-zag edges, and decays exponentially along the armchair direction. Spatial current patterns for a voltage bias applied (b) between the armchair edges, and (c) between the zig-zag edges.} \label{fig:edge}
\end{center}
\end{figure}

GNRs possess two low energy states near the middle of the band at $\pm E_l$, i.e., in close proximity to $E=0$, that are delocalized along the zig-zag edge, and localized along the direction of the armchair edge \cite{Oni08}. The form of charge transport through these localized states in the wide-lead limit has recently attracted some attention \cite{Mun06,Zar07,Tak10}, in particular in view of its possible application for DNA sequencing \cite{Saha12}. In figure~\ref{fig:edge}(a), we present the local density of states for a $(15 \times 7)$ GNR at $E=E_l=4.6 \times 10^{-13} t$, that clearly demonstrates the localized nature of these low energy states.
%
As a result, we expect that the charge transport at $V_c=E_l/e$ is highly anisotropic. To investigate this anisotropy, we consider the $(15 \times 7)$ GNR with four leads attached to sites along their zig-zag edges, as shown in figures~\ref{fig:edge}(b) and (c). When a voltage bias is applied between the armchair edges with $V_c=E_l/e$, (with chemical potential $\mu_L$ for leads 3 and 4, and $\mu_R$ for leads 1 and 2), the current flow is highly localized along the zig-zag edges, as shown in figure~\ref{fig:edge}(b), where the $E=E_l$ state is delocalized. In this case, the conductance of the $E=E_l$ state (in the four-lead configuration) is equal to the quantum of conductance (similarly, the state at $E=-E_l$ possesses a conductance given by the quantum of conductance). This result holds for all values of $N_a$ or $N_z$ that we have considered. Moreover, for an infinitely large GNR in the armchair direction, i.e., $N_a \rightarrow \infty$, one has a doubly degenerate state at $E_l=0$ (one localized state at each of the zig-zag edges), and the conductance of the GNR at $E=0$ is as expected equal to twice the quantum of conductance.
Note that the spatial current pattern in figure~\ref{fig:edge}(b) is qualitatively different from that obtained when wide leads are attached to the armchair edges: in this case, the largest current density occurs in the center of the GNR \cite{Mun06,Zar07,Tak10} and not along the zig-zag edges. On the other hand, when a voltage bias is applied between the zig-zag edges (with chemical potential $\mu_L$ for leads 1 and 3, and $\mu_R$ for leads 2 and 4), the current is strongly suppressed [see figure~\ref{fig:edge}(c)], and scales as $\sim E_l^{2}$. Since $E_{l}$ decreases exponentially with increasing length \(N_{a}\) \cite{Oni08}, we find that the current along the armchair direction also decreases exponentially with increasing length of the GNR, as expected from the localized nature of the $\pm E_{l}$-states.

\subsection{Current Patterns in Four Leads Configurations}
\label{sec:4leads}

Do quantum mechanical currents obey the principle of superposition? In particular, can spatial current patterns in GNRs that are attached to four leads, simply be obtained by superposing two-lead
current patterns such as the ones discussed above? To address this question we compare the spatial current patterns at $V_c=t/e$ in 4-lead and 2-lead configurations for two GNRs with different aspect ratios.

\begin{figure}
\begin{center}
\includegraphics[width=12cm]
{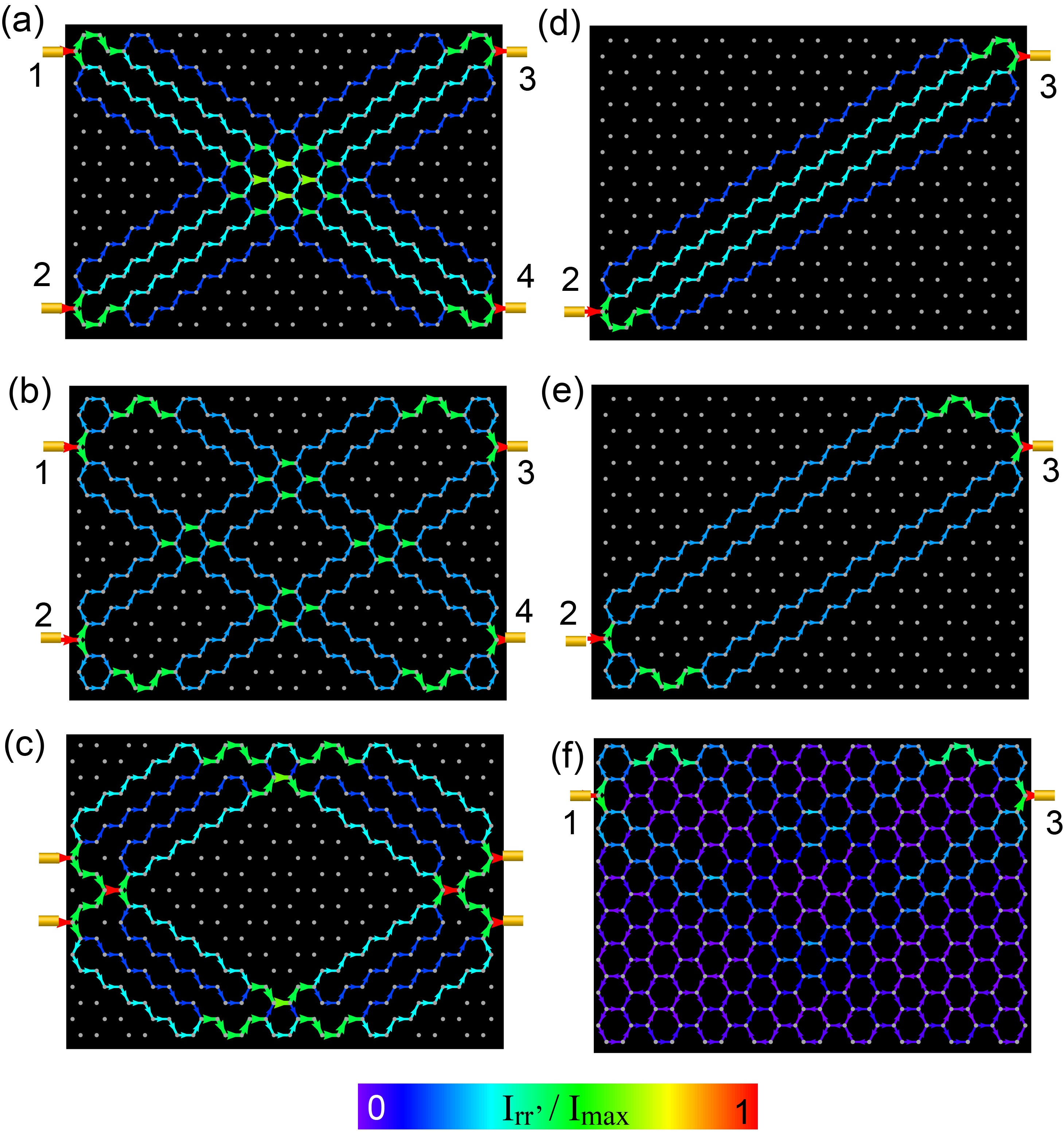} \caption{Comparison of spatial current patterns at $V_c=t/e$ in a $(9 \times 9)$ GNR for (a)-(c) 4-lead and (d) - (f) 2-lead configurations.} \label{fig:4Leads_1}
\end{center}
\end{figure}

In figure~\ref{fig:4Leads_1}, we present the current patterns for a $(9 \times 9)$ GNR in 4-lead configurations with the leads attached symmetrically around the center of the GNR, as shown in figures~\ref{fig:4Leads_1}(a) - (c), or with two leads attached, as shown in figures~\ref{fig:4Leads_1}(d) - (f).
Since the GNR possesses 9 degenerate states at $E=t$, we again find that the current follows predominantly the direction of the Fermi velocity, thus propagating along the direction of the primitive lattice vector. Moreover, changing the separation between the leads simply deforms the current pattern, but does not introduce any qualitatively new spatial elements. A comparison of the 4-lead current patterns in figures~\ref{fig:4Leads_1}(a) and (b) with the current patterns in a symmetric 2-lead configuration shown in figures~\ref{fig:4Leads_1}(d) and (e), respectively, demonstrates that the 4-lead current pattern indeed represents a superposition (i.e., is the sum) of two 2-lead current patterns. Moreover, the current flowing in the 4-lead configuration is twice that of the 2-lead configurations. This implies that in the 4-lead configuration, there are no interference effects between the two superposing 2-lead current patterns, and that, in particular, no current flow occurs between the upper (leads 1 and 3) or lower (leads 2 and 4) pair of leads. We also note that the 4-lead current pattern in figure~\ref{fig:4Leads_1}(b) is not the superposition of the 2-lead current pattern with just the upper two leads (leads 1 and 3)
in figure~\ref{fig:4Leads_1}(f), with its respective counterpart. In this case, the spatial current pattern is very diffuse, since the two leads cannot be connected by a
current path that propagates along ${\bf v}_F$, and the total current is only approximately 58.5\% of that in the 2-lead configuration of figure~\ref{fig:4Leads_1}(e).

\begin{figure}[h]
\begin{center}
\includegraphics[width=10cm]
{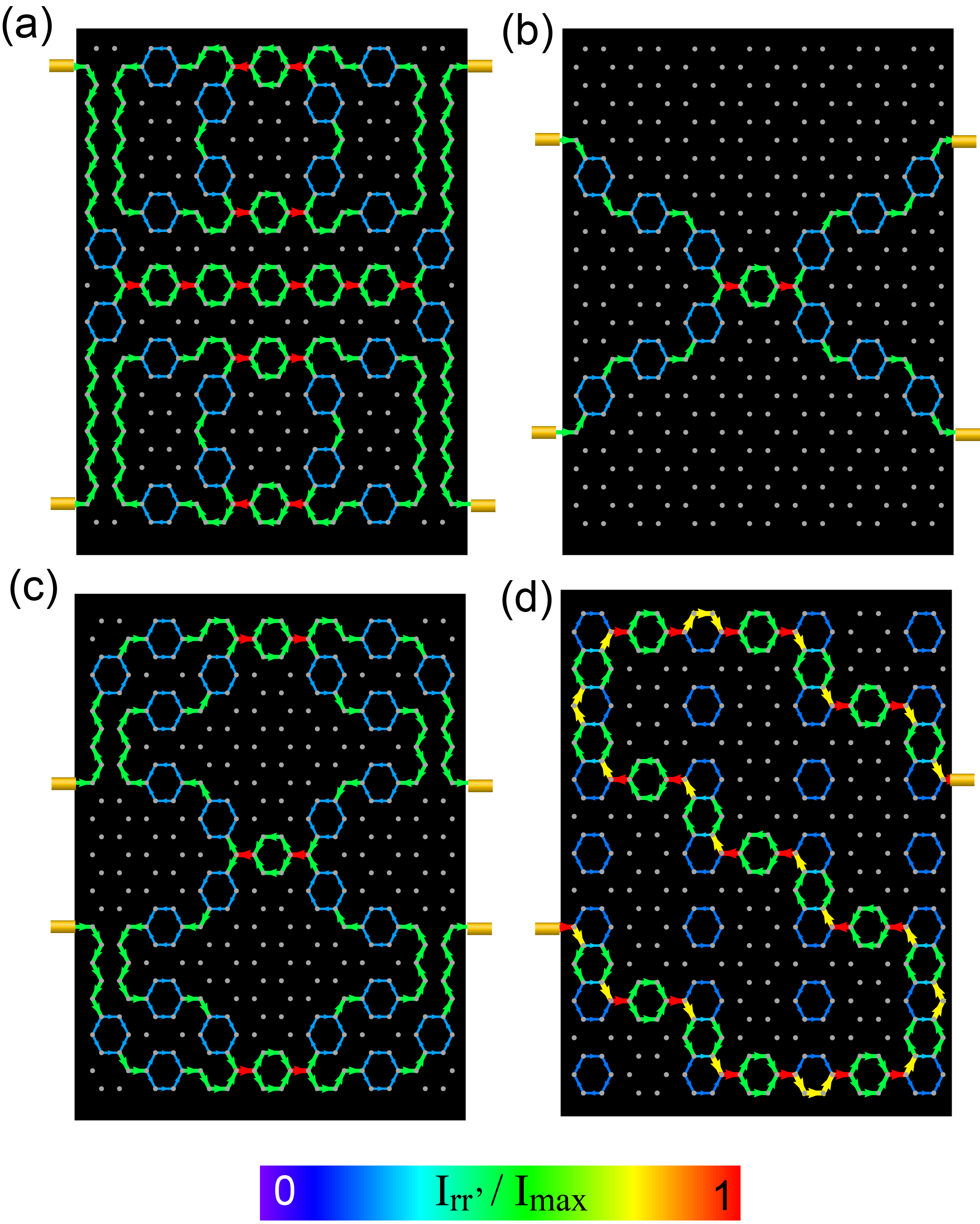} \caption{Comparison of spatial current patterns in a $(7 \times 13)$ GNR at $V_c=t/e$ for (a)-(c) 4-lead and (d) 2-lead configurations.} \label{fig:4Leads_2}
\end{center}
\end{figure}
In contrast, a $(7 \times 13)$ GNR possesses only a single state at $E=t$, and the spatial form and complexity of the current pattern at $V_c=t/e$ changes qualitatively when the separation between the leads on each side of the GNR is varied, as shown in figures~\ref{fig:4Leads_2}(a)-(c). When the leads are attached to the corners [see figure~\ref{fig:4Leads_2}(a)], the current flows along the edges and through the center of the GNR. At the same time, there are two large, circulating current loops in the upper and lower parts of the GNR, that do not contribute to the net current through the GNR. In contrast, when the separation between the leads is reduced [see figure~\ref{fig:4Leads_2}(b)], the current flows along the direction of ${\bf v}_F$, which directly connects the leads, while most of the GNR does not exhibit any charge flow at all. Upon reducing the separation between leads even further, the current pattern again becomes more complex [see figure~\ref{fig:4Leads_2}(c)]: the current now flows primarily along the outer perimeter, while connected to a larger circulating current loop in the center of the GNR. In figure~\ref{fig:4Leads_2}(d), we present the corresponding 2-lead current pattern; the superposition of this pattern with its counterpart yields a spatial current pattern that is identical to that of figure~\ref{fig:4Leads_2}(c). However, the total current through the 4-lead and 2-lead configurations is the same since [in contrast to the $(9 \times 9)$ GNR] there is only a single state that contributes to the charge transport at $V_c=t/e$, thus limiting the conductance for both configurations to a single quantum of conductance. In other words, the spatial current in the 4-lead configuration is half the sum of the currents of the two 2-lead configurations. The degeneracy of the $E=t$ state therefore does not only affect the spatial form of the current patterns in 2-lead configurations, but also whether the current patterns and total conductance in 4-lead configurations arises as a superposition of current patterns and conductances of 2-lead configurations.

\subsection{Diamond-shaped GNRs}

\begin{figure}
\begin{center}
\includegraphics[width=8cm]
{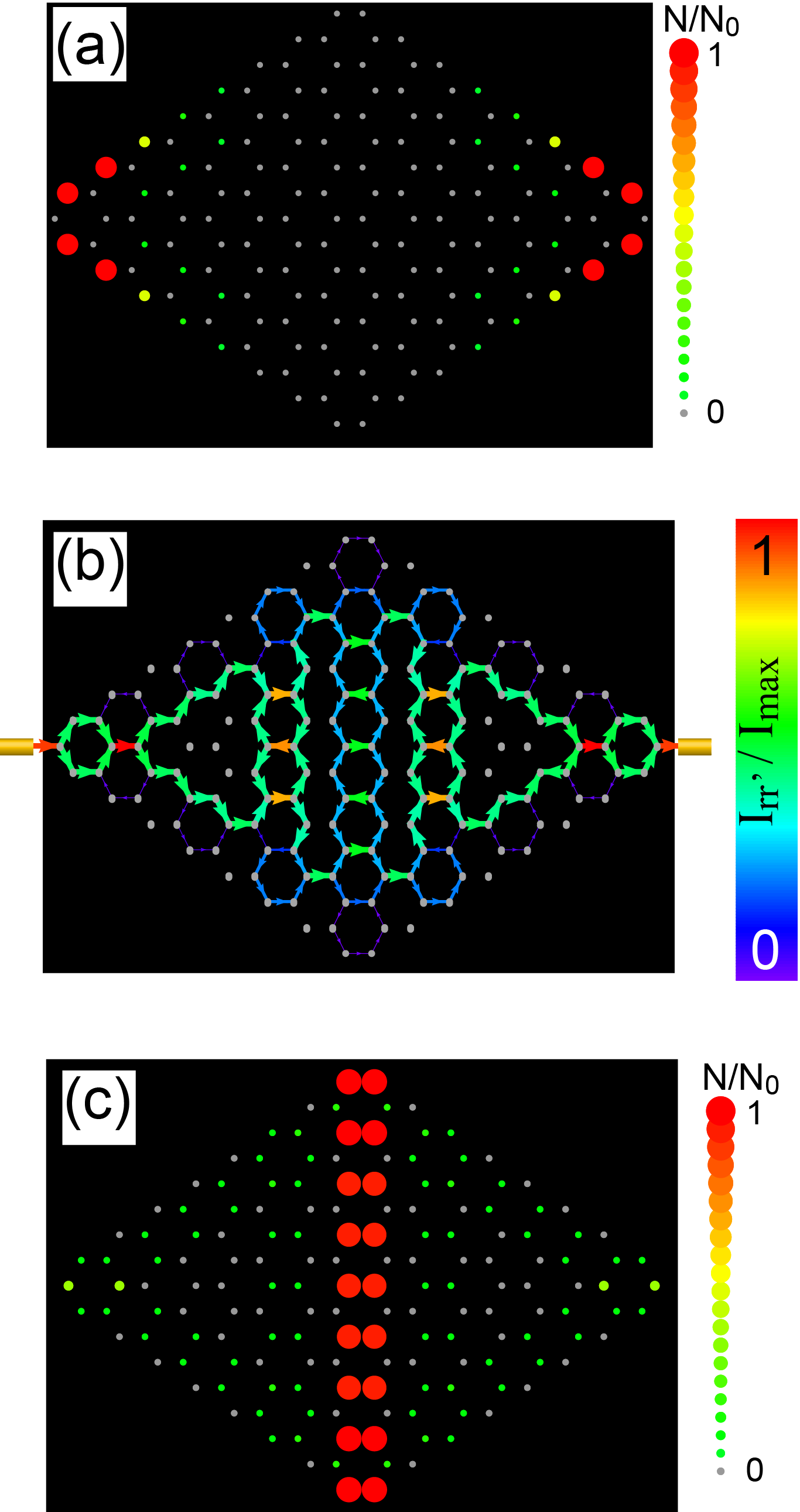} \caption{(a) Contour plot of the local density of states at $E=0$ in a diamond-shaped GNR. (b) Spatial current pattern for $V_c=t/e$, and (c) local density of states at $E=t$.} \label{fig:diamond}
\end{center}
\end{figure}

The observation that localized states exist near zig-zag edges in GNRs raises the interesting question of the form of localized state in other GNR geometries \cite{Mun06}. To investigate this question, we consider the diamond-shaped GNR shown in figure~\ref{fig:diamond} whose edges are only of the zig-zag type.
A plot of the local density of states at $E=0$ [see figure~\ref{fig:diamond}(a)] reveals that a localized state exists, but that it is not confined to the zig-zag edges of the GNR [as was the case for the GNR shown in figure~\ref{fig:edge}(a)] but to its two corners with acute angles. Moreover, while the GNR possesses three degenerate states at $E=t$, we find that the current does not follow the direction of the Fermi velocity (which would require it to flow along the edges of the GNR), but rather flows through the GNR's center, as shown in figure~\ref{fig:diamond}(b). Finally, a plot of the local density of states at $E=t$ in figure~\ref{fig:diamond}(c), which is the largest in the GNR's center, again reveals no resemblance to the spatial current pattern, as previously discussed in the context of figure~\ref{fig:x5}.

\subsection{Imaging of Spatial Current Patterns}
\label{sec:STM}

The ability to experimentally visualize spatial current patterns is crucial for understanding and manipulating transport at the nanoscopic or atomic scale. In mesoscopic systems, such as quantum point contacts \cite{Cro00,Top00_2,Top00,Jura07}, quantum rings \cite{Hac06,Mar07} and DNA \cite{Ter05}, imaging of spatial current paths was successfully achieved using scanning probe microscopy (SPM) \cite{reviews, Met05}. The spatial resolution of SPM, however, is insufficient to image spatial current patterns predicted to exist in nanoscopic systems \cite{Can12,Tod99,CLoops} which vary on the atomic scale. Two of us therefore recently proposed a novel method for the imaging of spatial current paths in nanoscopic systems based on scanning tunneling microscopy (STM) \cite{Can13}, a method that can resolve spatial current patterns on the scale of a lattice constant. In addition, in the experimentally realized weak tunneling limit \cite{Hof03}, this STM method only probes but essentially does not perturb the GNR's
electronic structure, in contrast to SPM.

To demonstrate that this method for the spatial imaging of currents can also successfully be applied to GNRs, we consider as an example the current patterns in figures~\ref{fig:7x}(a) and (e) for the $(7 \times 7)$ GNR, and in figure~\ref{fig:x5}(e) for the $(11 \times 5)$ GNR. To visualize these spatial current patterns, we plot the spatial dependence of the current, $I_{L,R}({\bf T})$, that flows from an STM tip (held above the GNR at potential $\mu_{T}$) through the GNR into the left (L) or right (R) lead as a function of tip position ${\bf T}$ (for a more detailed description, see Ref.~\cite{Can13}).
The tunneling of electrons from the STM tip into the GNR is described by the Hamiltonian in equation (\ref{eq:tunnel}). Here, we set the bias between the leads to zero, i.e., $\mu_{L,R} = \mu_{0}$, and require that the bias difference between the tip and the system, $\Delta V_T = (\mu_T - \mu_0) /e$ be equal to $\Delta V$, and that the bias midpoint $V^T_c = (\mu_0 + \mu_T)/2e$ be equal to $V_c$ (here, $V_c$ and $\Delta V$ were used to generate the actual current patterns).

\begin{figure}[h]
\begin{center}
\includegraphics[width=10.5cm]
{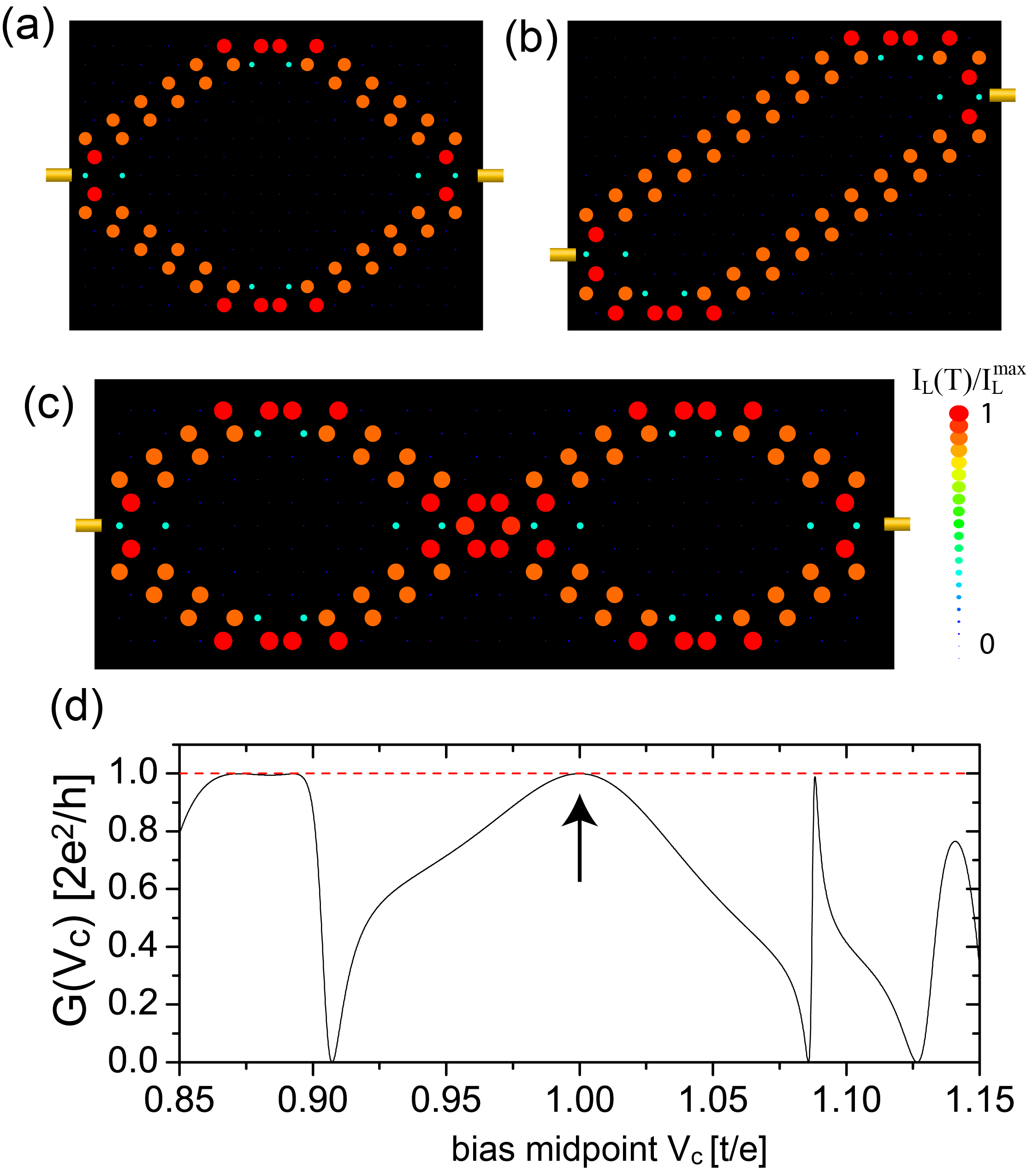} \caption{(a) - (c) Spatial plot of $I_{L}({\bf T})$ flowing from the STM tip into the left lead. (d) Conductance $G(V_c)$ as a function of bias midpoint $V_c$ for the $(7 \times 7)$ GNR. Here, we used $t_h=t$.} \label{fig:stm}
\end{center}
\end{figure}
In  figures~\ref{fig:stm}(a) - (c), we present the resulting STM image, i.e., the  spatial plots of $I_{L}({\bf T})$, which correspond to the current patterns in figures~\ref{fig:7x}(a) and (e) and in figure~\ref{fig:x5}(e), respectively. A comparison of the actual current patterns with the STM images shows that the latter provide an accurate and atomically resolved image of the spatial current patterns. This agreement holds even when the leads are not located at high symmetry positions, as shown in figures~\ref{fig:7x}(e) and \ref{fig:stm}(b), or when the size of the system increases, and the current pattern becomes repetitive, as in figures~\ref{fig:x5}(e) and \ref{fig:stm}(c).
Moreover, we demonstrated in Ref.~\cite{Can13}, that the form of the total conductance, $G(V_c)$, is an independent, experimentally verifiable, criterion for the success of the STM method in imaging spatial current patterns. In particular, we found that when $G(V_c)$ is close to the maximal allowed conductance, and varies weakly around a given $V_c$, the STM method successfully images the spatial current patterns at $V_c$ in the entire nanoscopic system. On the other hand, if $G(V_c)$ is sharply peaked, the agreement breaks down in at least part of the network. The $(7 \times 7)$ GNR considered here satisfies this criterion, as follows from figure~\ref{fig:stm}(d), where we present its conductance $G(V_c)$: it is close to the quantum of conductance at $V_c = t/e$ (see arrow) where we image the current pattern and its width (as a function of $V_c$) is larger than the distance between two peaks. Note that the zeros in the conductance $G(V_c)$ correspond to zeroes of the real part of the retarded Green's function for $t_h = 0$ which occur near, but never exactly on, conductance resonances. More generally, we find that the above criterion for $G(V_c)$ is satisfied in nanoscopic networks which (a) possess a continuum of energy eigenstates through which transport can take place or (b) are strongly coupled to leads.

Finally, we note in passing that understanding the magnitude and spatial paths of currents injected from the STM tip into the GNR, will likely also have relevance for DNA sequencing where currents are induced locally by pulling a DNA through a hole in a GNR \cite{Sch10,Min11,Ven11,Saha12}.

\subsection{Dephasing and the Classical Limit}

\begin{figure}
\begin{center}
\includegraphics[width=10cm]
{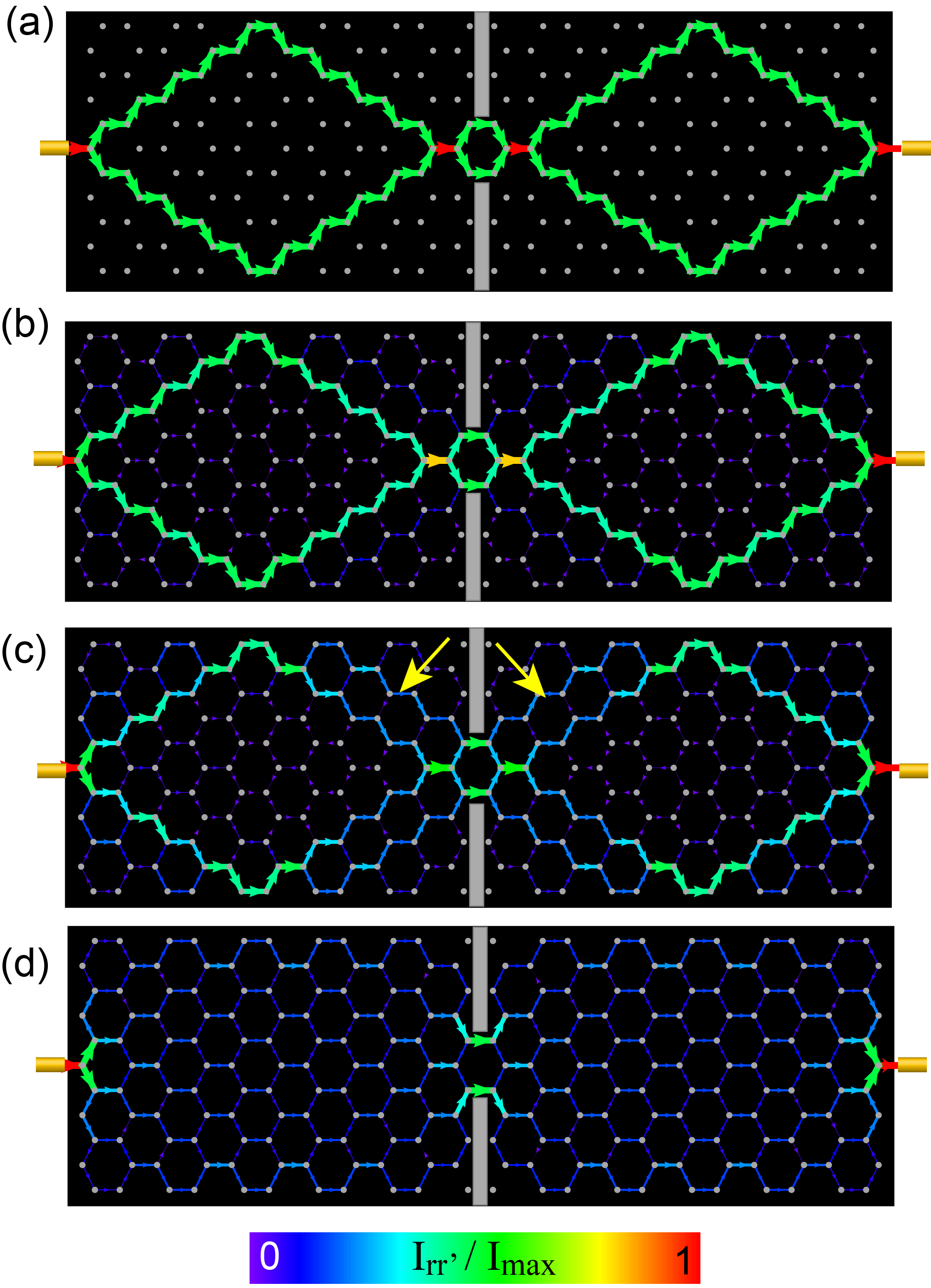} \caption{Evolution of spatial current patterns with increasing dephasing parameter, $\gamma$ for an $(11 \times 5)$ GNR with a constriction (shown in gray): (a) $\gamma= 0$ with mean free path $l=\infty$ (b) $\gamma= 0.005 t^2$, $l=11.5 a_0$, (c) $\gamma= 0.02 t^2$, $l=5.8 a_0$, (d) $\gamma= 0.5 t^2$, $l=1.2 a_0$.} \label{fig:dephasing}
\end{center}
\end{figure}
To study the effects of dephasing on the form of spatial current patterns, we consider the coupling of electrons to a local phonon mode \cite{Mon02,Gal07}, as described by the Hamiltonian in equation (\ref{eq:Hamiltonian}). As discussed in section~\ref{sec:theory}, we employ the high-temperature approximation, where the dephasing is controlled by a single parameter, $\gamma$ [see equation (\ref{eq:Sigma})]. In figure~\ref{fig:dephasing} we present the evolution of the spatial current pattern with increasing $\gamma$ for an $(11 \times 5)$ GNR that contains a constriction shown in gray. In the absence of dephasing, i.e., for $\gamma = 0$, [see figure~\ref{fig:dephasing}(a)], the current pattern is similar to that of the $(11 \times 5)$ GNR without a constriction, as shown in figure~\ref{fig:x5}(e). We note that while the constriction is located in a region of the GNR through which no current flows in figure~\ref{fig:x5}(e), it nevertheless leads to a change in the actual current pattern. As $\gamma$ increases, the current pattern becomes more diffuse, and evolves smoothly from the ballistic limit [see figure~\ref{fig:dephasing}(a)], to that of a classical resistor network [see figure\ref{fig:dephasing}(d)]. Moreover, with increasing $\gamma$, the current pattern remains well defined in the vicinity of the leads, but becomes more diffuse as one moves further away from the leads [see regions indicated by yellow arrows in figure~\ref{fig:dephasing}(c)]. This is expected since the diffuse current pattern arises from multiple scattering of the electrons off phonons, and the resulting dephasing. While required by symmetry, it is nevertheless interesting to note that the effects of dephasing are reversed (with the spatial current pattern again becoming more coherent) as the current approaches the site where it exits the GNR. This phenomenon of current patterns exhibiting coherent (well-defined) and incoherent (diffusive) spatial regions occurs when the mean-free path is considerably shorter than the size of the GNR [in figure~\ref{fig:dephasing}(c), one has  $l=5.8 a_0$], but larger than a few lattice spacings. As the mean-free path becomes shorter, the region around the leads where the current forms a well defined, coherent pattern shrinks and eventually vanishes [see figure~\ref{fig:dephasing}(d)].

\subsection{Effects of Next-Nearest Neighbor Hoppings on Current Patterns}
\label{sec:NNN}

In the preceding sections, we neglected the effects of a possible next-nearest neighbor hopping, $t^\prime$, on the form of the electronic excitation spectrum and the resulting conductance and spatial currents patterns in GNRs. The question naturally arises to what extent a non-zero $t^\prime$, with estimates ranging from $t^\prime=0.02t$ to $t^\prime=0.2t$ \cite{Rei02}, qualitatively affects the results presented above.

In macroscopic graphene lattices, the existence of a non-zero $t^\prime$ is often neglected, not only because of the large uncertainty in its value, but also because its effects on the electronic excitation spectrum are rather trivial. In particular, the energy dispersion for a non-zero $t^\prime$  is given by \cite{Wal47}
\begin{equation}
E_{\pm}({\bf k}) = \pm t \sqrt{3+f_{\bf k}} - t^\prime f_{\bf k}
\label{eq:Ektp}
\end{equation}
where
\begin{equation}
f_{\bf k} = 2 \cos(\sqrt{3} k_y a_0) + 4 \cos\left(\frac{\sqrt{3}}{2} k_y a_0\right)\cos\left(\frac{3}{2} k_x a_0\right)
\end{equation}
As a result, all momentum states located at an energy $E_0$ for $t^\prime =0$ are simply shifted to a new energy
\begin{equation}
E^\prime = E_0 - t^\prime \left[ \left( \frac{E_0}{t} \right)^2 -3 \right] \end{equation}
for $t^\prime \not =0$. This, in particular, implies that the degeneracy of states is not lifted by a non-zero $t^\prime$.

For finite-size GNRs, however, the situation is different since a non-zero $t^\prime$ lifts the degeneracy of the states which for $t^\prime=0$ are located at $E=t$ states. The extent of the energy splitting between these states is non-universal and depends not only on the aspect ratio of the GNR, but also on its overall size, since, according to Eq.(\ref{eq:Ektp}), the splitting vanishes in the limit of infinitely large graphene layers where degeneracy is restored.
\begin{figure}
\begin{center}
\includegraphics[width=15cm]
{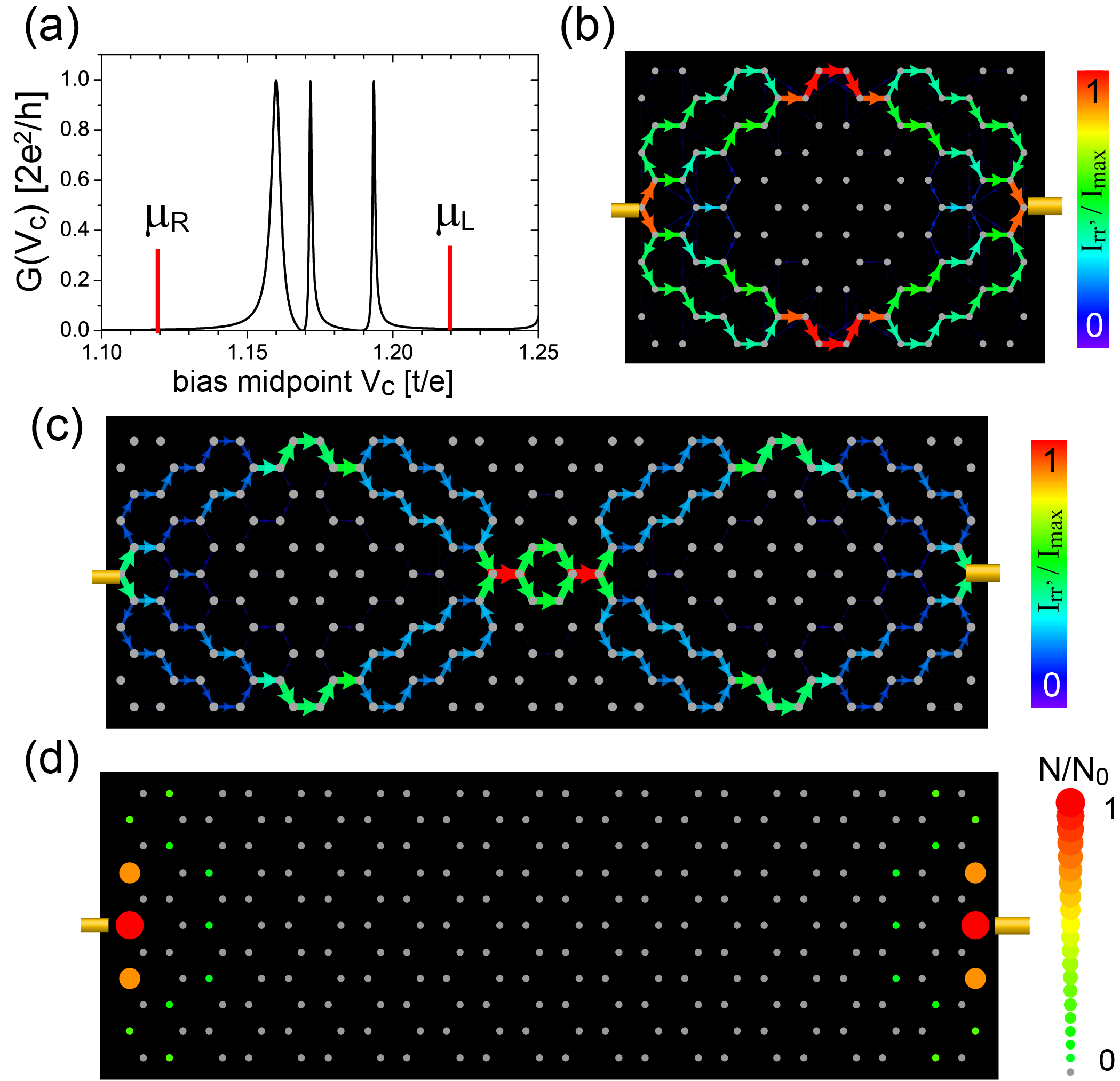} \caption{ (a) Conductance trace near the states which occur at \(E = t\) when \(t^\prime = 0\). If $\mu_{L,R}$ are chosen to bracket all three conductance resonances, well-ordered current patterns are produced for the (b) \((5 \times 5)\) GNR at \(t^\prime = 0.1t\) with \(\delta = 10^{-11} t\), \(\Delta \mu = 0.1t\), and \(V_{c} = 1.17 t/e\), and (c) \((11\times 5)\) GNR at \(t^\prime = 0.06t\) with \(\delta = 10^{-13} t\), \(\Delta \mu = 0.055 t\), and \(V_{c} = 1.1075 t/e\). (d) The LDOS for the edge states at energy \(E_l \approx 0.128 t\). } \label{fig:NNN}
\end{center}
\end{figure}
In figure \ref{fig:NNN}(a), we present a conductance scan for a $(5 \times 5)$ GNR with $t^\prime  = 0.1t$ over an energy range that contains all three states (and only those) that are located at $E=t$ for $t^\prime = 0$, and possess a non-zero wavefunction at ${\bf L}$ and ${\bf R}$. Choosing a left and right chemical potential that brackets these three states [see vertical red lines in figure \ref{fig:NNN}(a)], we find that the resulting spatial current pattern shown in figure \ref{fig:NNN}(b) still agrees very well with that obtained for the case $t^\prime = 0$, as follows from a comparison of figures \ref{fig:NNN}(b) and \ref{fig:x5}(b). The same conclusion also holds for larger GNRs and different values of $t^\prime$, as follows from a comparison of the current pattern for a $(11 \times 5)$ GNR with $t^\prime=0.06t$ shown in figure \ref{fig:NNN}(c) with that for  $t^\prime=0$ shown in figure \ref{fig:x5}(e). This implies that for both cases, $t^\prime = 0$ and $t^\prime \not = 0$,  the same (well-ordered) spatial current pattern occurs if the left and right chemical potentials are chosen such that they bracket all states which for $t^\prime = 0$ occur at $E=t$.

Finally, we found that a non-zero $t^\prime$ shifts the energy of the state localized along the zig-zag edge near $E \approx 0$ for $t^\prime = 0$, but does not change its localized nature or spatial structure. In figure \ref{fig:NNN}(d) we present the local density of states for $t^\prime  = 0.06t$ at $E_l \approx 0.128 t $, which demonstrates that this state remains localized along the zig-zag edge.

\section{Summary}

In this article, we investigated the quantum nature of local charge transport in graphene nanoribbons, as reflected in the form of spatial current patterns. By using the non-equilibrium Keldysh Green's function formalism \cite{theory,Car71a}, we demonstrated that the form of spatial current patterns is determined by the interplay between the GNRs' geometry, size and aspect ratio, by the location and number of leads, and the presence of dephasing. In particular, we identified a crucial relation between the spatial form of current patterns, and the number of degenerate states participating in the transport. This insight, in principle, provides us with the opportunity to predict and custom-design spatial current patterns in GNRs. Furthermore, we showed that the number of degenerate states participating in charge transport also determines whether current patterns and conductances in GNRs with 4-lead configurations can be considered as arising from the superposition of current patterns and conductances in 2-lead configurations. In addition, we showed that spatial current patterns in GNR can be spatially imaged with atomic resolution using scanning tunneling microscopy, allowing us to gain unique insight into the nature of local charge transport. We also demonstrated how spatial current patterns evolve with increasing dephasing from the ballistic limit, where they are coherent and spatially well defined, to the classical limit, where they are incoherent and spatially diffuse. Finally, we showed that our conclusions remain valid in the presence of a realistic next-nearest neighbor hooping element.  These results represent an important first step towards understanding and manipulating charge transport at the atomic level in GNRs, which are a necessary requirement for the further development of graphene based nanoelectronics and DNA sequencing.\\

\noindent {\bf Acknowledgements}

We would like to thank  D. Goldhaber-Gordon and H. Manoharan  for
stimulating discussions. This work is supported by the U.S. Department of Energy, Office of Basic Energy Sciences, Division of Materials Sciences and Engineering under Award No.~DE-FG02-05ER46225 (D.K.M) and by a U.S. Department of Education GAANN Fellowship and MRSEC NSF DMR-0820054 (T.C.).

\end{document}